\documentstyle[aps,epsf,twocolumn]{revtex}
\tightenlines
\begin{document}
\draft
\title{
Magnon Exchange Mechanism of Ferromagnetic Superconductivity
         }
\author{Naoum Karchev\cite{byline}}
\address{
Department of Physics, University of Sofia, 1126 Sofia, Bulgaria
      }
\maketitle
\begin{abstract}

The magnon exchange mechanism of ferromagnetic superconductivity (FM-superconductivity) 
was developed to explain in
a natural way the fact that the 
superconductivity in $UGe_2$, $ZrZn_2$ and 
$URhGe$ is confined to the ferromagnetic 
phase.The order parameter is a spin anti-parallel component of 
a spin-1 triplet with zero spin projection. The transverse spin 
fluctuations are pair forming and the longitudinal ones are pair 
breaking. In the present paper, a superconducting solution, based on the
magnon exchange mechanism, is obtained which closely matches the experiments
with $ZrZn_2$ and $URhGe$. The onset of superconductivity leads to the appearance of
complicated Fermi surfaces in the spin up and spin down momentum 
distribution functions. Each of them consist of two pieces, but they are
simple-connected and can be made very small by varying the microscopic parameters.
As a result, it is obtained that the specific heat depends on the temperature
linearly, at low temperature, and the coefficient $\gamma=\frac {C}{T}$ is
smaller in the superconducting phase than in the ferromagnetic one.
The absence of a quantum transition from ferromagnetism to ferromagnetic
superconductivity in a weak ferromagnets $ZrZn_2$ and $URhGe$ is explained accounting for
the contribution of magnon self-interaction to the spin fluctuations' parameters. It is
shown that in the presence of an external magnetic field the system undergoes
a first order quantum phase transition. 
  
\end{abstract}


{\bf I Introduction}
\ \\

Very recently ferromagnetic superconductivity (FM-superconductivity) 
has been observed 
in $UGe_2$\cite{me1}, $ZrZn_2$\cite{me2} and $URhGe$\cite{me3}.
The superconductivity is confined to the ferromagnetic phase. 
Ferromagnetism and superconductivity are believed to arise due to the same 
band electrons. The persistence of ferromagnetic order
within the superconducting phase has been ascertained by neutron scattering.
The specific heat anomaly associated with the superconducting transition 
in these materials appears to be absent.

At ambient pressure $UGe_2$ is an itinerant ferromagnet below the Curie
temperature
 $T_c=52K$, with low-temperature ordered moment of $\mu_s=1.4\mu_B/U$. 
With increasing pressure the system passes through two
successive quantum phase transition, from ferromagnetism to
FM-superconductivity at $P\sim 10$ kbar, and at higher pressure $P_c\sim$ 16
kbar to paramagnetism\cite{me1,me4}.
At the pressure where the superconducting 
transition temperature is a maximum  
$T_{sc}=0.8K$, 
the ferromagnetic state is still stable with $T_c=32K$, and 
the system undergoes a first order metamagnetic transition between two
ferromagnetic phases with different ordered moments \cite{rp1}.  
The specific heat coefficient 
$\gamma=\frac {C}{T}$ increases
steeply near 11 kbar and retains a large and nearly 
constant value\cite{me5}.  

The ferromagnets $ZrZn_2$ and $URhGe$ are superconducting at ambient pressure 
with superconducting critical temperatures $T_{sc}=0.29K$ and $T_{sc}=0.25K$ 
respectively.
$ZrZn_2$ is ferromagnetic below the Curie temperature $T_c=28.5K$ with 
low-temperature ordered moment of $\mu_s=0.17\mu_B$ per formula unit, while 
for $URhGe$\,\,
$T_c=9.5K$
and $\mu_s=0.42\mu_B$. The low Curie temperatures and small ordered moments
indicate that compounds are close to a ferromagnetic quantum critical point.
A large jump in the specific heat, at the temperature where the resistivity
becomes zero, is observed in $URhGe$. At low temperature the specific
heat coefficient $\gamma$ is twice smaller than in the ferromagnetic phase.
materials 

The most popular theory of FM-superconductivity is based on the paramagnon
exchange mechanism\cite{me6,me7}. The order parameters are spin
parallel components of the spin triplet. The superconductivity in
$ZrZn_2$ was predicted, but the theory meets many difficulties. 
In order to explain the absence of superconductivity in paramagnetic phase 
it was accounted for the magnon
paramagnon interaction and proved that the critical temperature is much
higher in the ferromagnetic phase than in the paramagnetic one\cite{me8}.
To the same purpose, the Ginzburg-Landau mean-field theory was modified
with an exchange-type interaction between the magnetic moments of 
triplet-state Cooper pairs and the ferromagnetic 
magnetization density\cite{me9}.

The Fay and Appel (FA) theory predicts that
spin up and spin down fermions
form Cooper pairs, and hence the specific heat
decreases
exponentially at low temperature. The phenomenological theories
\cite{mm9} circumvent the problem assuming that only
majority spin fermions form pairs, and hence the minority spin
fermions contribute the asymptotic of the specific heat. The coefficient
$\gamma=\frac CT$ is twice smaller in the
superconducting phase, which closely matches the experiments with
$URhGe$\cite{me3}
but does not resemble the experimental results for $UGe_2$ and $ZrZn_2$.
The assumption seems to be doubtful for systems with very small 
ordered moment. 

The superconducting critical temperature in
(FA) theory increases when the magnetization decreases and very close to the
quantum critical point falls down rapidly.  It has recently been the subject
of controversial debate. It is obtained in\cite{mm11}, by means of a more
complete Eliashberg treatment, that the transition temperature is nonzero at
the critical point. In \cite{mm12},  however, the authors have shown that the
reduction of quasiparticle coherence and life-time due to spin fluctuations is
the pair-breaking  process which leads to a rapid reduction of the
superconducting critical temperature near the quantum critical point. 

Recent studies of polycrystalline samples of $UGe_2$ show that T-P phase
diagram
 is very similar to those of single-crystal specimens of $UGe_2$
\cite{mm13}. These
 findings suggest that the superconductivity in $UGe_2$ is
relatively 
 insensitive to the presence of impurities and defects which
excludes
 the spin parallel pairing.

Despite of the efforts, the improved theory of paramagnon induced
superconductivity can not cover the whole variety of properties of 
FM-superconductivity. 

Magnon exchange mechanism of superconductivity has
been proposed \cite{rp2} to explain in 
a natural way the fact that the  
superconductivity in $UGe_2$, $ZrZn_2$ and  
$URhGe$ is confined to the ferromagnetic  
phase.The order parameter is a spin anti-parallel component  
$\uparrow\downarrow+\downarrow\uparrow$ of  
a spin-1 triplet  $(\uparrow\uparrow, 
\uparrow\downarrow+\downarrow\uparrow, \downarrow\downarrow)$ 
with zero spin projection. The transverse spin  
fluctuations are pair forming and the longitudinal ones are pair  
breaking. The competition between magnons and paramagnons explains 
the existence of the two successive quantum phase transitions in 
$UGe_2$.

An itinerant system is considered in which the
spin-$\frac 12$ fermions responsible for the ferromagnetism are the same
quasiparticles which form the Cooper pairs. 
Hence, one has to consider the equation for the gap 
as well as the equation for the magnetization.
Then the system of equations for the gap and for the magnetization determines 
the phase where the superconductivity and the ferromagnetism coexist.

In the present paper, a superconducting solution, based on the
magnon exchange mechanism, is obtained which closely matches the experiments
with $ZrZn_2$ and $URhGe$. The proposed superconducting solution differs from 
the superconducting solution discussed in\cite{rp2} in two ways.
First, the quantum phase transition from ferromagnetism to 
FM-superconductivity is smooth second order phase transition, which resembles the 
experimental results for $UGe_2$. In the present paper, 
the system passes trough a first order quantum phase
transition. Second, to form a Cooper pair an
electron transfers from one Fermi surface to the other. As a result, the
onset of superconductivity in \cite{rp2} leads to the appearance of two Fermi
surfaces in each of the spin up and spin down momentum distribution functions. 
The existence of the two Fermi surfaces explains the linear dependence
of the specific heat at low temperatures as opposed to the exponential
decrease of the specific heat in the BCS theory. In the ferromagnetic phase
the specific heat constant $\gamma$  is smaller than in the superconducting one, 
which closely matches the experiments with $UGe_2$\cite{me5}. 
The onset of superconductivity, in the preset paper, leads to the appearance of
complicated Fermi surfaces in the spin up and spin down momentum 
distribution functions. Each of them consist of two pieces, but they are
simple-connected and can be made very small by varying the microscopic parameters. 
As a result, $\gamma=\frac {C}{T}$ can be made smaller in FM-superconducting phase in 
agreement with $URhGe$ experiments\cite{me3}.  
  
The existence of two Fermi surfaces in each of the spin-up and spin-down
momentum distribution functions is a generic property of a 
FM-superconductivity with spin anti-parallel pairing. An important example is 
the coexistence of ferromagnetism and s-superconductivity induced by phonons 
\cite{mm15}. The spin fluctuations are pair breaking if the order parameter is
spin singlet \cite{mm13a}, and superconductivity and ferromagnetism coexist if 
the spin fluctuations are weak, while the magnon-induced superconductivity coexist
with ferromagnetism close to quantum critical point where the spin fluctuations are
very strong. The present paper and \cite{mm15} describe different physical realities
which lead to coexistence of ferromagnetism and superconductivity.

The paper is organized as follows. In Sec.II an effective Hamiltonian is obtained. 
In Sec III, the magnon-induced superconductivity is discussed. The superconducting 
solution known from \cite{rp2} is reported to complete the investigation. 
Section IV is devoted to the concluding remarks.

\ \\

{\bf II Effective Hamiltonian}
\ \\

An itinerant system is considered in which the  
spin-$\frac 12$ fermions responsible for the ferromagnetism are the same  
quasiparticles which form the Cooper pairs. 
The effective interaction of quasiparticles  
$c_{\sigma}(\vec x)(c^+_{\sigma}(\vec x))$ with spin fluctuations has the form 
\FL 
\begin{equation}   
H_{\text{s-f}}=J\int d^3x \,c^+(\vec x)\frac {\vec \tau}{2}c(\vec x)\cdot   
\vec {M}(\vec x) 
\label{mm1a}   
\end{equation}  
where the transverse spin fluctuations are described by  
magnons  
\FL 
\begin{eqnarray} 
M_1(\vec x)+iM_2(\vec x) & = & \sqrt{2M}a(\vec x),\nonumber \\ 
M_1(\vec x)-iM_2(\vec x) & = & \sqrt{2M}a^+(\vec x)
\label{mm1b} 
\end{eqnarray} 
and the longitudinal spin fluctuations by paramagnons 
\FL 
\begin{equation} 
M_3(\vec x)-M=\varphi(\vec x). 
\label{mm1c} 
\end{equation} 
$M$ is zero temperature dimensionless magnetization of the system per 
lattice site.  
 
The partition function  can be written as a path integral over the complex  
functions of the Matsubara time $\tau$ \,\, 
$a(\tau,\vec x),a^+(\tau,\vec x),\varphi(\tau,\vec x)$ and Grassmann
functions  $c_{\sigma}(\tau,\vec x),c^+_{\sigma}(\tau,\vec x)$ 
\cite{pi} 
\FL
\begin{equation}
{\cal Z}(\beta)\,=\,\int D\mu\left(a^+,a,\varphi,c^+_{\sigma},c_{\sigma}\right)
e^{-S}.
\label{2mm1}
\end{equation}
 
The action is a
sum of free action $S_0$ and part which describes the spin-fermion
interaction $S_{\text{int}}$.  
\FL
\begin{eqnarray}
S_0 & = & \int\limits^{\beta}_0 d\tau\int \frac {d^3k}{(2\pi)^3}
\left[a^+(\tau,\vec k)\dot a(\tau, \vec k)+
\omega(\vec k)a^+(\tau,\vec k)a(\tau,\vec k)\right. \nonumber \\
& & \left.+\,\,
\varphi(\tau,\vec k)D^{-1}_{pm}(\tau,\vec k)\varphi(\tau,-\vec k)\right. \\
\label{2mm2}
& & \left.+\,\, c^+_{\sigma}(\tau,\vec k)\dot c_{\sigma}(\tau, \vec k)
+\epsilon_{\sigma}(\vec k) c^+_{\sigma}(\tau,\vec k)c_{\sigma}(\tau,\vec k)
\right] \nonumber
\end{eqnarray}
where $\beta$ is the inverse temperature.  
The magnon's dispersion is  
\FL 
\begin{equation} 
\omega(\vec k)=\rho\vec k^2 
\label{magnon}  
\end{equation} 
where the spin stiffness  
constant is proportional to $M$ ($\rho=M\rho_0$). The paramagnon  
propagator in Matsubara representation is 
\FL  
\begin{equation}   
D_{\text{pm}}(\tau,\vec k)=\int\frac {d\omega}{(2\pi)} 
\frac {e^{i\omega \tau}}{r+\frac {|\omega|}{|\vec k|}+b\vec k^2}.  
\label{para}  
\end{equation}   
The parameter $r$ is the inverse static longitudinal magnetic susceptibility,  
which measures the deviation from quantum critical point. The constants  
$J,\rho_0$ and $b$ are phenomenological ones subject to some relations.  
Finally, the spin-up and spin-down fermions have the following dispersion relations: 
\FL  
\begin{equation}   
\epsilon_{\uparrow}(\vec k)= \frac {\vec k^2}{2m}-\mu-\frac {JM}{2},  
\quad \,\,\,
\epsilon_{\downarrow}(\vec k)=\frac {\vec k^2}{2m}-\mu +\frac {JM}{2}  
\label{disp}  
\end{equation}   
Accounting for Eqs(\ref{mm1b}) and (\ref{mm1c}) one obtains the following
expression for   the interacting part of the action 
\FL 
\begin{eqnarray} 
S_{\text{int}} & = & \frac {J}{2}\int\limits^{\beta}_0d\tau\int d^3x 
\left[\sqrt {2M} c_{\uparrow}^+(\tau,\vec x)c_{\downarrow}(\tau,\vec
x)a(\tau,\vec x)\right. \nonumber \\
& & \left.  +\,\sqrt {2M} c_{\downarrow}^+(\tau,\vec x)
c_{\uparrow}(\tau,\vec x)a^+(\tau,\vec x)\right. \\
\label{2mm3}
& & \left.+\,
\left(c_{\uparrow}^+(\tau,\vec x)c_{\uparrow}(\tau,\vec x)- 
c_{\downarrow}^+(\tau,\vec x)c_{\downarrow}(\tau,\vec
x)\right)\varphi(\tau,\vec x)\right]   \nonumber
\end{eqnarray} 
 
The integral Eq.(\ref{2mm1}) over the bose fields ($a,a^+,\varphi$) is
Gaussian. Integrating them out, using the formula for the Gaussian integral 
\cite{pi}, 
one obtains a representation 
for the partition function in terms of path integral over the Grassmann fields.  
\FL
\begin{equation}
{\cal Z}(\beta)\,=\,\int D\mu\left(c^+,c\right) e^{-S_{eff}}.
\label{2mm4}
\end{equation}

The effective fermion action
$S_{eff}$ is a sum of free part and
resulting four-fermion interaction $S_{f^4}$  
\FL 
\begin{eqnarray} 
& & S_{f^4}\,=\,-\frac {J^2}{8}\int
d^4x_1d^4x_2\left[c_{\uparrow}^+(x_1)c_{\uparrow}(x_1)
- c_{\downarrow}^+(x_1)c_{\downarrow}(x_1)\right] \nonumber \\
& & D_{pm}(x_1-x_2) 
\left[c_{\uparrow}^+(x_2)c_{\uparrow}(x_2)-c_{\downarrow}^+(x_2)c_{\downarrow}(x_2)\right]\,- \\
\label{2mm5} 
& & \frac {MJ^2}{2}\int d^4x_1d^4x_2
c_{\downarrow}^+(x_1)c_{\uparrow}(x_1)D_{m}(x_1-x_2) 
c_{\uparrow}^+(x_2)c_{\downarrow}(x_2). \nonumber
\end{eqnarray} 
where $x=(\tau,\vec x)$, $D_{\text{pm}}$ is the paramagnon propagator Eq.(\ref{para}), and $D_{\text{m}}$ 
is the magnon Green function 
\FL
\begin{equation}
D_{m}(x)\,=\,\int\frac {d\omega}{2\pi}\frac {d^3k}{(2\pi)^3}
\frac {e^{-i\omega\tau+i\vec k \vec x}}{i\omega+\rho\vec k^2}
\label{2mm6}
\end{equation}
 
For the purpose of doing analytical calculations it is convenient to approximate the four-fermion 
interaction with the static one. To this end I replace the magnon and paramagnon propagators 
Eqs.(\ref{2mm6},\ref{para}) by static potentials 
\FL
\begin{eqnarray}
2M\,D_{m}(\omega,\vec k)\rightarrow & & V_{m}(\vec k)=\frac {2M}{\rho\vec k^2} \\
\label{pot}
D_{pm}(\omega,\vec k)\rightarrow & & V_{pm}(\vec k)=\frac {1}{r+b\vec k^2}.
\nonumber
\end{eqnarray} 
 
The next step is to represent the spin anti-parallel composite field 
$c_{\uparrow}c_{\downarrow}$ as a sum of symmetric and antisymmetric parts. 
After some algebra one obtains an effective four 
fermion theory which can be written as a sum of four terms. Three of them 
describe the interaction of the components of spin-1 composite fields 
$(\uparrow\uparrow,
\uparrow\downarrow+\downarrow\uparrow,\downarrow\downarrow)$   which have a
projection of spin 1,0 and -1 respectively.   The fourth term describes the
interaction of the spin singlet composite fields  
$\uparrow\downarrow-\downarrow\uparrow$. The 
Hamiltonians of interactions are  
\FL  
\begin{eqnarray}  
H_{\uparrow\uparrow} & = & -\frac {J^2}{8}\int\prod\limits_{i}\frac  
{d^3k_i}{(2\pi)^3}  
\left[c_{\uparrow}^+(\vec k_1)c_{\uparrow}^+(\vec k_2)\right. \nonumber \\  
& & \left. c_{\uparrow}(\vec k_2-\vec k_3)c_{\uparrow}(\vec k_1+\vec 
k_3)\right]  
 V_{pm}(\vec k_3) 
\label{up}  
\end{eqnarray} 
\FL 
\begin{eqnarray}  
H_{\downarrow\downarrow} & = & -\frac {J^2}{8}\int\prod\limits_{i}\frac  
{d^3k_i}{(2\pi)^3}  
\left[c_{\downarrow}^+(\vec k_1)c_{\downarrow}^+(\vec k_2)\right. \nonumber \\ 
& & \left. c_{\downarrow}(\vec k_2-\vec k_3)c_{\downarrow}(\vec k_1+\vec 
k_3)\right]  
 V_{pm}(\vec k_3) 
\label{down}  
\end{eqnarray} 
\FL 
\begin{eqnarray}  
& & H_{p} = -\frac  
{J^2}{8}\int\prod\limits_{i}\frac  {d^3k_i}{(2\pi)^3}  
\left[c_{\uparrow}^+(\vec k_1)c_{\downarrow}^+(\vec k_2)+  
c_{\downarrow}^+(\vec k_1)c_{\uparrow}^+(\vec k_2)\right] \nonumber\\ 
\label{uds} 
\\ 
& & \left[c_{\uparrow}(\vec k_2-\vec k_3)c_{\downarrow}(\vec k_1+\vec k_3)+  
c_{\downarrow}(\vec k_2-\vec k_3)c_{\uparrow}(\vec k_1+\vec k_3)\right]  
V_{-}(\vec k_3) \nonumber 
\end{eqnarray} 
\FL 
\begin{eqnarray}  
& & H_{s} = \frac  
{J^2}{16}\int\prod\limits_{i}\frac  {d^3k_i}{(2\pi)^3}  
\left[c_{\uparrow}^+(\vec k_1)c_{\downarrow}^+(\vec k_2)-  
c_{\downarrow}^+(\vec k_1)c_{\uparrow}^+(\vec k_2)\right] \nonumber \\  
\label{udas} 
\\ 
& & \left[c_{\uparrow}(\vec k_2-\vec k_3)c_{\downarrow}(\vec k_1+\vec k_3)-  
c_{\downarrow}(\vec k_2-\vec k_3)c_{\uparrow}(\vec k_1+\vec k_3)\right]  
V_{+}(\vec k_3) 
\nonumber 
\end{eqnarray}  
where  
\FL 
\begin{equation}  
V_{-}(\vec k)=\frac {2M}{\rho\vec k^2}\,-\,\frac {1}{r+b\vec k^2},\qquad   
V_{+}(\vec k)=\frac {2M}{\rho\vec k^2}\,+\,\frac {1}{r+b\vec k^2}   
\label{pot2}  
\end{equation}  
 
The spin singlet fields' interaction Eq.(\ref{udas}) 
is repulsive and does not contribute to the superconductivity\cite{mm13a}.  
The spin parallel fields' interactions Eqs.(\ref{up},\ref{down}) are 
due to the exchange of paramagnons and  do not contribute to the 
magnon-mediated superconductivity. The relevant  
interaction is that of the $\uparrow\downarrow+\downarrow\uparrow$ fields 
Eq.(\ref{uds}). It has an attracting part due to exchange of magnons and a 
repulsive part due to  exchange of paramagnons.  
 
The effective Hamiltonian of the system is 
\FL   
\begin{equation}  
H_{\text eff}=H_0+H_{p}  
\label{effh}  
\end{equation}  
where  $H_0$ is the Hamiltonian of the free spin  
up and spin down fermions with dispersions Eq.(\ref{disp}). 
\ \\

{\bf III Magnon-Induced Superconductivity}
\ \\

By means of the Hubbard-Stratanovich transformation one introduces  
$\uparrow\downarrow+\downarrow\uparrow$ composite field and then the fermions 
can be integrated out. The obtained free energy is a function 
of the composite field and the integral over the composite field can be 
performed approximately by means of the steepest descend method. To this end 
one sets the first derivative of the free energy with respect to composite 
field equal to zero, this is the gap equation. To ensure that the 
fermions which form Cooper pairs are the same as those responsible for 
spontaneous magnetization, one has to consider the equation for the 
magnetization as well.
\FL
\begin{equation}  
M=\frac 12 <c^+_{\uparrow}c_{\uparrow}-c^+_{\downarrow}c_{\downarrow}> 
\label{mag} 
\end{equation} 
The system of equations for the gap and for the magnetization
determines  the phase where the superconductivity and the ferromagnetism
coexist.

The system can be written in terms of Bogoliubov excitations, which have 
the following dispersions relations:
\FL 
\begin{eqnarray} 
& & E_1(\vec k) = -\frac 
{JM}{2}-\sqrt{\epsilon^2(\vec k)+|\Delta(\vec 
k)|^2} \nonumber \\  
& & E_2(\vec k) = \frac 
{JM}{2}-\sqrt{\epsilon^2(\vec 
k)+|\Delta(\vec k)|^2}   
\label{bog} 
\end{eqnarray} 
where $\Delta(\vec k)$ 
is the gap, and $\epsilon(\vec k)=\frac {\vec k^2}{2m}-\mu$.  
At zero temperature the equations take the form 
\FL
\begin{eqnarray} 
M & = & \frac 12\int\frac {d^3k}{(2\pi)^3}\left[1-\Theta (-E_2(\vec k))\right] 
\label{eqs1}\\ 
\Delta(\vec p) & = & \frac {J^2}{8}\int\frac {d^3k}{(2\pi)^3}\,\frac {V(\vec 
p-\vec k)\,\,\Theta (-E_2(\vec k))}{\sqrt {\epsilon^2(\vec k)+|\Delta(\vec 
k)|^2}}\,\Delta (\vec k) 
\label{eqs2} 
\end{eqnarray} 
 
The gap is an antisymmetric function $\Delta (-\vec k)=-\Delta (\vec k)$, so 
that the expansion in terms of spherical harmonics $Y_{lm}(\Omega_{\vec k})$ 
contains only terms with odd $l$. I assume that the component with $l=1$ and 
$m=0$ is nonzero and the other ones are zero
\FL  
\begin{equation} 
\Delta (\vec k)=\Delta_{10}(k)\sqrt {\frac {3}{4\pi}}\cos\theta. 
\label{gap} 
\end{equation} 
Expending the potential $V_{-}(k)$ in terms of Legendre polynomial $P_l$ 
one obtains that only the 
component with $l=1$ contributes the gap equation. The potential $V_1(p,k)$ 
has the form,
\FL 
\begin{eqnarray} 
V_{1}(p,k) & = & \frac {3M}{\rho}\left[\frac 
{p^2+k^2}{4p^2k^2}\ln\left(\frac {p+k}{p-k}\right)^2-\frac{1}{pk}\right] 
- \nonumber \\ 
& &  
\frac {3M}{\rho}\beta \left[\frac {p^2+k^2}{4p^2k^2}\ln\frac 
{r'+(p+k)^2}{r'+(p-k)^2}\,-\,\frac {1}{pk}\right],  
\label{pot3} 
\end{eqnarray} 
where $\frac {3M}{\rho}=\frac {3}{\rho_0}$, $\beta=\frac {\rho}{2Mb}=\frac 
{\rho_0}{2b}>1$ and $r'=\frac {r}{b}<<1$. A straightforward analysis  
shows that for a fixed $p$ , the potential is positive when $k$ 
runs an interval around $p$ $(p-\Lambda,p+\Lambda)$, 
where $\Lambda$ is approximately independent on $p$.  
In order to allow for an explicit analytic solution, I introduce further 
simplifying assumptions by neglecting the dependence of $\Delta_{10}(k)$ 
on $k$ ($\Delta_{10}(k)=\Delta_{10}(p_f)=\Delta$) and setting 
$V_1(p_f,k)$ equal to a constant $V_1$ within interval 
$(p_f-\Lambda,p_f+\Lambda)$ and zero elsewhere. The system of equations 
(\ref{eqs1},\ref{eqs2})  
is then reduced to the system
\FL 
\begin{eqnarray} 
M & = & \frac 
{1}{8\pi^2}\int\limits_0^{\infty}dkk^2\int\limits_{-1}^{1}dt[1-\Theta(-E_2(k,t))] 
\label{eqsb1}\\ 
\Delta & = & \frac {J^2V_1}{32\pi^2}\int\limits_{p_f-\Lambda}^{p_f+\Lambda} dk 
k^2\int\limits_{-1}^{1}dt\,t^2\frac 
{\Theta(-E_2(k,t))}{\sqrt{\epsilon^2(k)+\frac {3}{4\pi}t^2\Delta^2}} 
\Delta 
\label{eqsb2} 
\end{eqnarray} 
where $t=\cos\theta$.

\ \\ 

{\bf  III.1 Solution which satisfies $\sqrt {\frac {3}{\pi}}\Delta<JM$}
\ \\

The equation of magnetization 
(\ref{eqsb1}) shows that it is convenient to represent the gap in the form  
$\Delta= \sqrt {\frac {\pi}{3}}\kappa (M) JM$, where $\kappa (M)<1$. 
Then the equation
\FL
\begin{equation} 
E_2(k,t)=0, 
\label{mm8a}
\end{equation}
defines the Fermi surfaces,
\FL 
\begin{equation} 
p^{\pm}_{f}=\sqrt {p^2_{f}\pm m\sqrt{J^2M^2-\frac 
{3}{\pi}t^2\Delta^2}}\,\,,\,\,\, p_{f}=\sqrt {2\mu m} 
\label{mm8} 
\end{equation}  

The domain between the Fermi surfaces contributes to the magnetization $M$ in 
Eq.(\ref{eqsb1}), but it is cut out from the domain of integration in the gap 
equation Eq.(\ref{eqsb2}). When the magnetization increases, the 
domain of integration in the gap equation decreases. Near the quantum 
critical point the size of the gap is small, and hence the linearized gap 
equation can be considered. Then it is easy to obtain the critical value of 
the magnetization $M_{SC}$ \cite{rp2}
 
When the magnetization approaches 
zero, the domain between the Fermi surfaces decreases. One can approximate the 
equation for magnetization Eq.(\ref{eqsb1}) substituting $p^{\pm}_{f}$ from 
Eq.(\ref{mm8}) in the 
the difference $(p^{+}_{f})^2-(p^{-}_{f})^2$ and setting 
$p^{\pm}_{f}=p_{f}$ elsewhere. Then, in this approximation, the 
magnetization is linear in $\Delta$, namely 
\FL
\begin{equation} 
\Delta =\sqrt {\frac {\pi}{3}}J\kappa M 
\label{mm10} 
\end{equation} 
where $\kappa$ runs the interval $(0,1)$, and satisfies the equation
\FL 
\begin{equation} 
\kappa\sqrt {1-\kappa^2}+\arcsin\kappa\,=\frac {8\pi^2}{mp_{f}J} 
\label{mm11} 
\end{equation} 
The Eq.(\ref{mm11}) has a solution if $mp_{f}J>16\pi$. Substituting $M$ from 
Eq.(\ref{mm10}) in Eq.(\ref{eqsb2}), one arrives at an equation for the gap.
This  equation can be solved in a standard way and the solution is
\FL 
\begin{eqnarray} 
\Delta & = & \sqrt {\frac {16\pi}{3}}\frac {\Lambda p_{f}\kappa}{m} 
\exp\left[-\frac32 
I(\kappa)-\frac {24\pi^2}{J^2 V_1 m p_{f}}\right] 
\label{mm12} \\ 
I(\kappa) & = & \int\limits_{-1}^{1}dt t^2 \ln\left(1+\sqrt 
{1-\kappa^2 t^2}\right) \nonumber 
\end{eqnarray} 
Eqs (\ref{mm10},\ref{mm11},\ref{mm12}) are the solution of the system 
Eqs.(\ref{eqsb1},\ref{eqsb2}) near the quantum transition to paramagnetism.  

One can write the momentum 
distribution functions $n^{\uparrow}(p,t)$ and $n^{\downarrow}(p,t)$ of the 
spin-up and spin-down quasiparticles 
in terms of the distribution functions of the Bogoliubov fermions  
\FL
\begin{eqnarray} 
n^{\uparrow}(p,t) & = & u^2(p,t)n_1(p,t)+v^2(p,t)n_2(p,t) 
\label{mm13} \\ 
n^{\downarrow}(p,t) & = & u^2(p,t)(1-n_1(p,t))+v^2(p,t)(1-n_2(p,t)) 
\nonumber 
\end{eqnarray} 
where $u(p,t)$ and $v(p,t)$ are the coefficients in the Bogoliubov 
transformation. At zero temperature $n_1(p,t)=1$, 
$n_2(p,t)=\Theta(-E_2(p,t))$, and the Fermi surfaces Eq.(\ref{mm8}) manifest 
themselves both in the spin-up and spin-down momentum distribution functions. 
The functions are depicted in Fig.1 and Fig.2.

The two Fermi surfaces explain the mechanism of Cooper pairing.   
In the ferromagnetic phase $n^{\uparrow}$ and $n^{\downarrow}$ have different  
(majority and minority) Fermi surfaces (see Fig.3 and Fig.4, $t=0$ graphs).  
The spin-up electrons contribute the majority 
Fermi surface, and spin-down electrons contribute the minority Fermi surface. 
When the value of the momentum of the emitted or absorbed magnon lies within  
interval $(p_f-\Lambda,p_f+\Lambda)$ the effective potential between spin-up and  
spin-down electrons is attracting. Hence, if the Fermi momenta $p_f^{\uparrow}$ and 
$p_f^{\downarrow}$ lie within interval $(p_f-\Lambda,p_f+\Lambda)$ the interaction 
between spin-up electrons, which contribute the majority Fermi surface, and spin-down  
electrons, which contribute the minority Fermi surface, is attracting. As a result,  
spin-up electrons from majority Fermi surface transfer to the minority Fermi surface 
and form spin anti-parallel Cooper pairs, while spin-down electrons from minority 
Fermi surface transfer to the majority one and form spin anti-parallel Cooper pairs too. 
As a result, the onset of superconductivity is accompanied by the appearance of a  
second Fermi surface in each of the spin-up and spin-down momentum  
distribution functions (see Fig.3 and Fig.4, $t=1$ graphs). 

\begin{figure}[h]  
\vspace{0.03cm}  
\epsfxsize=8.0cm  
\hspace*{0.2cm}  
\epsfbox{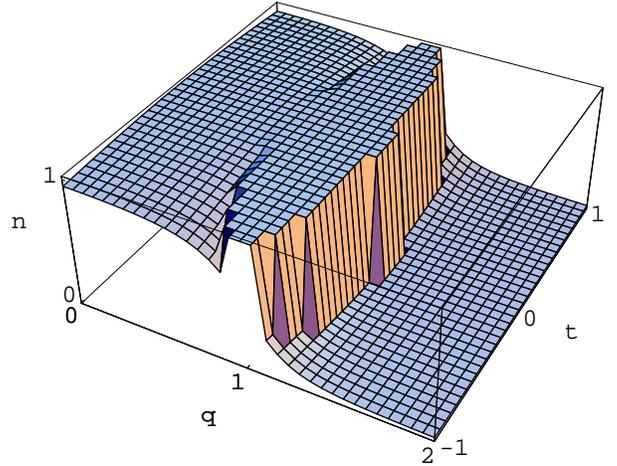}  
\caption{The zero temperature momentum distribution $n$, for spin up  
fermions, as a function of $q=\frac {p}{p_{f}}$ and $t=\cos\theta$.}  
\label{fig1}  
\end{figure}  
 
\begin{figure}[h]  
\vspace{0.03cm}  
\epsfxsize=8.0cm  
\hspace*{0.2cm}  
\epsfbox{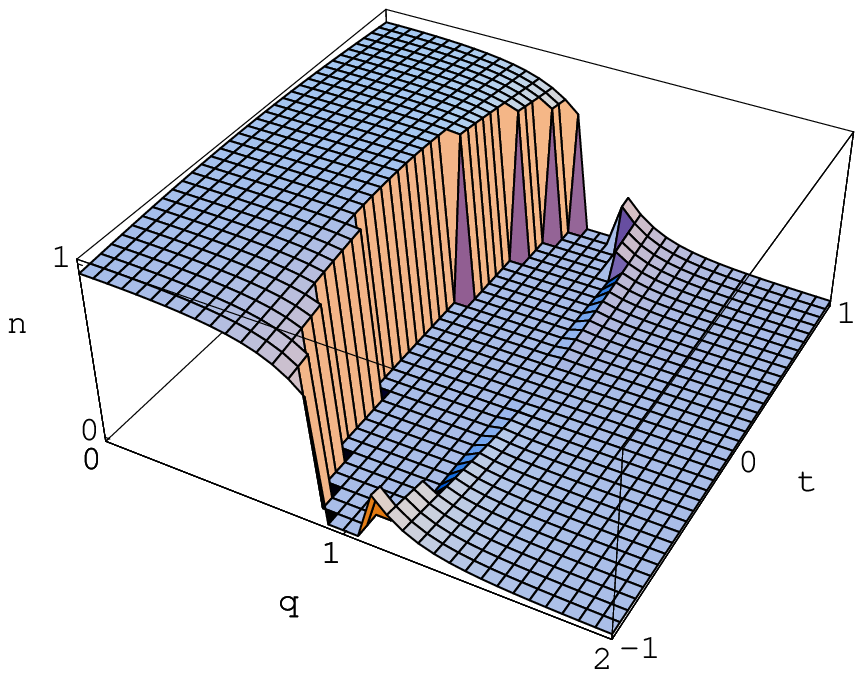}  
\caption{The zero temperature momentum distribution $n$, for spin down 
fermions, as a function of $q=\frac {p}{p_{f}}$ and $t=\cos\theta$.}  
\label{fig2}  
\end{figure}

\begin{figure}[h]  
\vspace{0.3cm}  
\epsfxsize=8.0cm  
\hspace*{0.2cm}  
\epsfbox{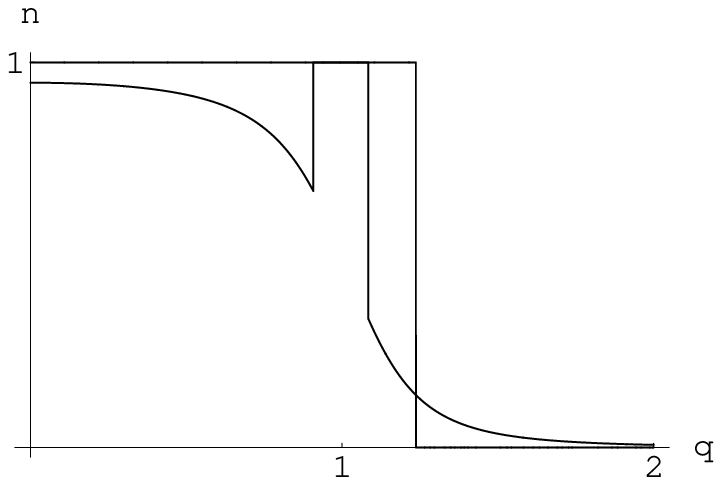}  
\caption{The zero temperature momentum distribution $n$, for spin up  
fermions, as a function of $q=\frac {p}{p_{f}}$ for $t=0$
(the gap iz zero) 
and $t=\pm 1$ (the gap is maximal).}  
\label{fig3}  
\end{figure}  
 
\begin{figure}[h]  
\vspace{0.3cm}  
\epsfxsize=8.0cm  
\hspace*{0.2cm}  
\epsfbox{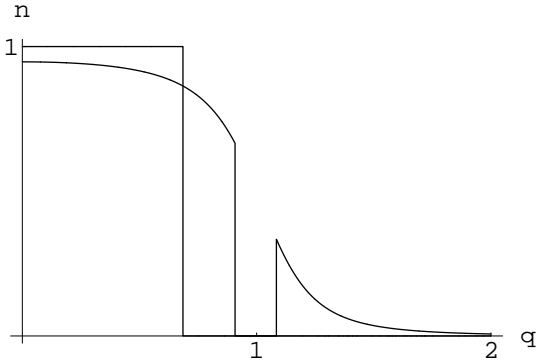}  
\caption{The zero temperature momentum distribution $n$, for spin down 
fermions, as a function of $q=\frac {p}{p_{f}}$ for $t=0$ (the gap is zero) 
and $t=\pm 1$ (the gap is maximal).}  
\label{fig4}  
\end{figure} 

The existence of the two Fermi surfaces explains the linear dependence 
of the specific heat at low temperatures:
\FL 
\begin{equation} 
\frac {C}{T}\,=\,\frac {2\pi^2}{3}\left(N^+(0)+N^-(0)\right) 
\label{mm14} 
\end{equation} 
Here $N^{\pm}(0)$ are the density of states on the Fermi surfaces. 
One ca rewrite the $\gamma=\frac {C}{T}$ constant in terms of Elliptic 
Integral of the second kind $E(\alpha,x)$ 
\FL
\begin{eqnarray} 
\gamma\,=\,\frac {m p_{f}}{3\kappa} & & 
\left[(1+s)^{\frac 12} E(\frac 12 \arcsin\kappa,\frac 
{2s}{s+1})+\right. \nonumber  \\   
& & \left. (1-s)^{\frac 12} E(\frac 12 \arcsin\kappa,\frac 
{2s}{s-1})\right]. 
\label{mm15} 
\end{eqnarray} 
where $s=\frac {JMm}{p^2_{f}}<1$ and  
$\kappa=\sqrt {\frac {3}{\pi}}\frac {\Delta}{JM}$. 
The Eq.(\ref{mm15}) shows that in the ferromagnetic phase 
($\Delta=0$) the specific heat constant $\gamma$  is smaller than in 
the superconducting one, which closely matches the 
experiments with $UGe_2$.

\ \\

{\bf III.2 Solution which satisfies $\sqrt {\frac {3}{\pi}}\Delta>JM$} 
\ \\

In the present sub-chapter one
looks for a solution of the system which satisfies
\FL
\begin{equation}
\sqrt {\frac {3}{\pi}}\Delta\,>\,JM
\label{me7}
\end{equation} 
The inequality Eq.(\ref{me7})
shows that the gap can not be arbitrarily small when the magnetization
is finite. Hence the system undergoes the quantum phase transition from
ferromagnetism to FM-superconductivity with a jump. Approaching the quantum
critical point from the ferromagnetic side, one sets the gap equal to zero
in the equation for the magnetization (\ref{eqsb1}) and considers the gap 
equation (\ref{eqsb2}) with magnetization as a parameter. It is more 
convenient to consider the free energy as a function of the gap for the
different values of the parameter $M$. To this purpose I introduce the
dimensionless "gap" $x$ and the parameters $s,\lambda$ and $g$ 
\FL
\begin{equation} 
x=\sqrt {\frac {3}{\pi}}\frac {m}{p_{f}^2}\Delta,\,\, 
s=\frac {m}{p_{f}^2}JM,\,\,\lambda=\frac {\Lambda}{p_{f}},\,\, 
g=\frac {J^2V_1mp_{f}}{8\pi^2}  
\label{me8} 
\end{equation} 
Then the free energy is a function of $x$ 
and depends on the parameters $s, \lambda$ and $g$.
\FL 
\begin{eqnarray} 
& & F(x)=\frac {6m^2}{\pi p_{f}^4}\left({\cal F}(x)-{\cal F}(0)\right) = x^2+  
g\int\limits_{1-\lambda}^{1+\lambda}dq q^2\int\limits_{-1}^1dt\,\times\nonumber 
\\ & & \left[\left(s-\sqrt {(q^2-1)^2+t^2x^2}\right) 
\Theta(\sqrt{(q^2-1)^2+t^2x^2}-s)-\right. \nonumber \\ 
& & \left. \left(s-\sqrt {(q^2-1)^2}\right) 
\Theta(\sqrt{(q^2-1)^2}-s)\right] 
\label{me9} 
\end{eqnarray} 
The dimensionless free energy $F(x)$ is depicted in Fig.5 for
$\lambda=0.08,\,g=20$ and three values of the parameter $s$, $s=0.8, s=0.69$
and $s_{cr}=0.595$. As the graph shows, for some values of the microscopic
parameters $\lambda$ and $g$, and decreasing the parameter $s$ (the
magnetization), the system passes trough a first order quantum phase
transition. The critical values $s_{cr}$ and $x_{cr}$ satisfy $\frac
{x_{cr}}{s_{cr}}=\sqrt {\frac {3}{\pi}}\frac {\Delta_{cr}}{JM_{cr}}>1$ in
agreement with Eq.(\ref{me7}).
\begin{figure}[h]   
\vspace{0.0cm}   
\epsfxsize=8.0cm   
\hspace*{-0.15cm}   
\epsfbox{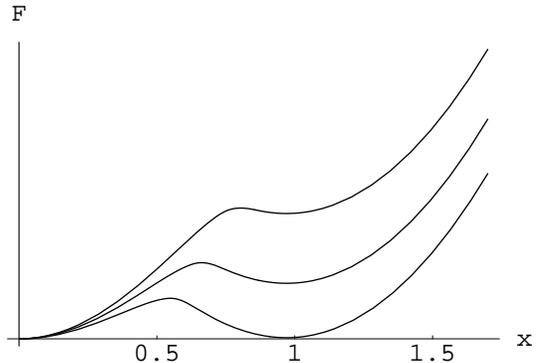}    
\caption{The dimensionless free energy $F(x)$ as a function of 
dimensionless gap $x$. $\lambda=0.08,\,g=20$, $s_1=0.8$(upper line), 
$s_2=0.69$(middle line) and 
 $s_{cr}=0.595$(lower line).}      
\end{figure}

Varying the microscopic parameters beyond the critical values, one has to solve 
the system of equations (\ref{eqsb1},\ref{eqsb2}). One represents again the gap in the form
\FL
\begin{equation}  
\Delta= \sqrt {\frac {\pi}{3}}\kappa (M) JM
\label{memfs1}
\end{equation} 
but now $\kappa (M)>1$.
Then the equation $E_2(k,t)=0$, which defines the Fermi surface, has no 
solution if 
$-1<t<-\frac {1}{\kappa (M)}$ and $\frac {1}{\kappa(M)}<t<1$, 
and has two solutions
\FL 
\begin{equation} 
p^{\pm}_{f}=\sqrt {p^2_{f}\pm m\sqrt{J^2M^2-\frac 
{3}{\pi}t^2\Delta^2}} 
\label{me10} 
\end{equation}  
when $-\frac {1}{\kappa(M)}<t<\frac {1}{\kappa (M)}$. 

The solutions (\ref{me10}) determine the two pieces of the Fermi surface. They stick 
together at $t=\pm\frac {1}{\kappa(M)}$, so that the Fermi surface is simple 
connected. The domain between pieces contributes to the 
magnetization $M$ in Eq.(\ref{eqsb1}), but it is cut out from the domain of 
integration in the gap equation Eq.(\ref{eqsb2}). The Fermi surface manifests itself 
both in the spin-up and spin-down momentum distribution functions. 
The functions are depicted in Fig.6 and Fig.7.

When the magnetization approaches  
zero, one can approximate the  
equation for magnetization Eq.(\ref{eqsb1}) substituting $p^{\pm}_{f}$ from  
Eq.(\ref{me10}) in the  
the difference $(p^{+}_{f})^2-(p^{-}_{f})^2$ and setting  
$p^{\pm}_{f}=p_{f}$ elsewhere. Then, in this approximation, the  
magnetization is linear in $\Delta$, namely  
\FL 
\begin{equation}  
\Delta =\sqrt {\frac {\pi}{3}}J\kappa M  
\label{me11}  
\end{equation}  
where $\kappa=\frac {mp_{f}J}{16\pi}$ is the small magnetization limit of  
$\kappa(M)$.  
The Eq.(\ref{me11}) is a solution if $mp_{f}J>16\pi$ (see Eq.(\ref{me7})).  
Substituting $M$ from  
Eq.(\ref{me11}) in Eq.(\ref{eqsb2}), one arrives at an  
equation for the gap. This 
equation can be solved in a standard way and the 
solution is 
\FL 
\begin{equation}  
\Delta\,=\,\sqrt {\frac {16\pi}{3}}  
\frac {p_{f}\Lambda}{m}  
\exp \left[-\frac {24\pi^2}{mp_{f}J^2V_1}-\frac {\pi}{4\kappa^3}+\frac  
{1}{3}\right]  
\label{me12}  
\end{equation}  
Eqs (\ref{me11},\ref{me12}) are the solution of the system  
near the quantum transition to paramagnetism.   
The second derivative of the free energy Eq.(\ref{me9}) with respect  
to the gap is positive when $\frac {mp_{f}J}{16\pi}>(\frac {21\pi}{16})^{\frac  
13}$,  hence the state where the  
superconductivity and the ferromagnetism coexist is stable. 

The existence of the Fermi surface explains the linear dependence  
of the specific heat at low temperature:   
\begin{equation}  
\frac {C}{T}\,=\,\frac {2\pi^2}{3}N(0)  
\label{me14}  
\end{equation}  
Here $N(0)$ is the density of states on the Fermi surface.  
One can rewrite the $\gamma=\frac {C}{T}$ constant in terms of Elliptic  
Integral of the second kind $E(\alpha,x)$  
\begin{eqnarray}  
\gamma\,=\,\frac {m p_{f}}{3\kappa(M)} & &  
\left[(1+s)^{\frac 12} E(\frac {\pi}{4},\frac  
{2s}{s+1})+\right. \nonumber  \\    
& & \left. (1-s)^{\frac 12} E(\frac {\pi}{4},\frac  
{2s}{s-1})\right].  
\label{me15}  
\end{eqnarray}  
where $s<1$ (see Eq.(\ref{me8})).  

\begin{figure}[h]  
\vspace{0.03cm}  
\epsfxsize=8.0cm  
\hspace*{0.0cm}  
\epsfbox{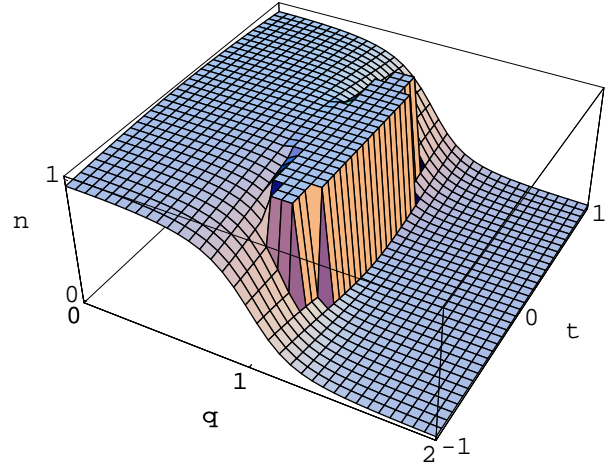}  
\caption{The zero temperature momentum distribution $n$, for spin up  
fermions, as a function of $q=\frac {p}{p_{f}}$ and $t=\cos\theta$.}  
\label{fig2}  
\end{figure}  
 
\begin{figure}[h]  
\vspace{0.3cm}  
\epsfxsize=8.0cm  
\hspace*{0.2cm}  
\epsfbox{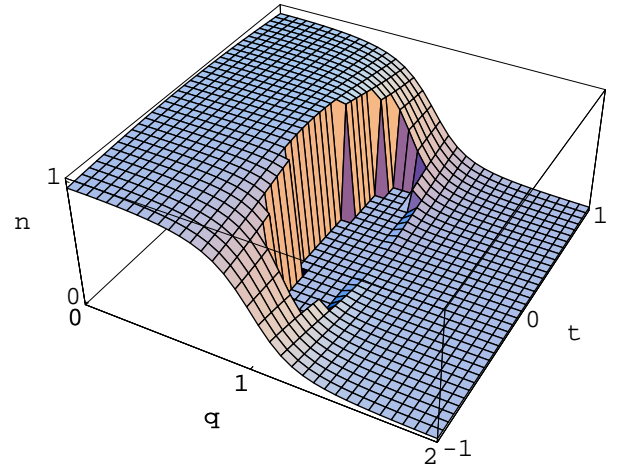}  
\caption{The zero temperature momentum distribution $n$, for spin down 
fermions, as a function of $q=\frac {p}{p_{f}}$ and $t=\cos\theta$.}  
\label{fig3}  
\end{figure} 

Eq.(\ref{me15}) shows that for $\kappa(M)$ just above one the specific heat 
constant $\gamma$  is smaller in ferromagnetic phase, while for $\kappa(M)>>1$
it is smaller in FM-superconducting phase.The result closely matches the 
experiments with $ZrZn_2$ and $URhGe$ respectively.  

The solutions Eqs.(\ref{mm10},\ref{me11})show that magnetization and
superconductivity disappear simultaneously. It results from the equation of
magnetization, which in turn is added to ensure that the 
fermions which form Cooper pairs are the same as those responsible for 
spontaneous magnetization. Hence, the fundamental assumption that
superconductivity and ferromagnetism are caused by the same electrons leads
to the experimentally observable fact that the quantum phase transition is a
transition to paramagnetic phase without superconductivity.

An important experimental fact is that $ZrZn_2$ and $URhGe$ are
superconductors at ambient pressure as opposed to the existence of a quantum
phase transition in $UGe_2$. To comprehend this difference one
considers the potential (\ref{pot3}). The quantum phase transition results from
the existence of a momentum cutoff $\Lambda$, above which the potential is
repulsive. In turn, the cutoff excistence follows from the relation
$\beta=\frac {\rho}{2Mb}>1$, which is true when the spin-wave approximation
expression for the spin stiffness constant $\rho=M\rho_0$ is used. The spin
wave approximation correctly describes systems with a large magnetization, for
example $UGe_2$. But in order to study systems with small magnetization, one
has to account for the magnon-magnon interaction which changes the small
magnetization  asymptotic of $\rho$, $\rho=M^{1+\alpha}\rho_0$, where
$\alpha>0$. Then for a small $M$ $\beta<1$, and the potential is attractive
for all momenta. Hence for systems which, at ambient pressure, are close to
quantum critical point, as $ZrZn_2$ and $URhGe$, the magnon self-interaction
renormalizes the spin fluctuations parameters so that the magnons dominate
the pair formation and quantum phase transition can not be observed. But if
one applies an external magnetic field, the magnon opens a gap proportional
to the magnetic field. Increasing the magnetic field the paramagnon
domination leads to first order quantum phase transition.  

\ \\

{\bf IV Conclusions}
\ \\

The proposed model of ferromagnetic superconductivity differs from the
models discussed in \cite{me6,me7,me8,me9} in many aspects. First, the
superconductivity is due to the exchange of magnons, and the model describes
in an unified way the superconductivity in $UGe_2,\,ZrZn_2$ and $URhGe$.
Second, the paramagnons have
pair-breaking effect. So, the understanding the mechanism of paramagnon
suppression is crucial in the search for the ferromagnetic superconductivity
with higher critical temperature. For example, one can build such a bilayer
compound that the spins in the two layers are oriented in two
non-collinear directions, and the net ferromagnetic moment is nonzero.
The paramagnon in this phase is totally suppressed and the low lying excitations
consist of magnons and additional spin wave modes with linear dispersion
$\epsilon(k)\sim k$\cite{me13}. If the new spin-waves are pair breaking, their
effect is weaker than those of the paramagnons, and hence the superconducting
critical temperature should be higher. 
Third, the order parameter is a
spin antiparallel component of a spin triplet with zero spin projection.
The existence of two Fermi surfaces in each of the spin-up and spin-down
momentum distribution functions 
leads to a linear temperature dependence of the specific heat at low
temperature.

The proposed model of magnon-induced superconductivity does not contain the
relativistic effects, namely spin-orbital coupling which is present in $UGe_2$.
The resulting magneto-crystalline anisotropy will modify the spin-wave
excitation and will add a gap in the magnon spectrum, which changes the
potential Eq.(\ref{pot3}). The physical consequence of the change is that the
superconductivity disappears before the quantum phase transition from
ferromagnetism to paramagnetism (see \cite{me5,mm13}). The distance between
these two points depends on the anisotropy. 

\ \\

\acknowledgments

The author would like to thank C. Pfleiderer for valuable discussions.

\end{document}